\documentclass[aps,prl,reprint]{revtex4-1}
\usepackage{placeins}
\usepackage{graphicx}
\usepackage{amssymb, amsmath}

\begin{document}

\title{Model-Independent Inference of Neutron Star Radii from Moment of Inertia Measurements}
\author{Carolyn A. Raithel, Feryal \"Ozel, \& Dimitrios Psaltis}
\affiliation{Department of Astronomy and Steward Observatory, University of Arizona, 933 N. Cherry Avenue, Tucson, Arizona 85721, USA}

\begin{abstract}
A precise moment of inertia measurement for PSR J0737$-$3039A in the double pulsar system is expected within the next five years. We present here a new method of mapping the anticipated measurement of the moment of inertia directly into the neutron star structure. We determine the maximum and minimum values possible for the moment of inertia of a neutron star of a given radius based on physical stability arguments, assuming knowledge of the equation of state only at densities below the nuclear saturation density. If the equation of state is trusted up to the nuclear saturation density, we find that a measurement of the moment of inertia will place absolute bounds on the radius of PSR~J0737$-$3039A to within $\pm$1 km. The resulting combination of moment of inertia, mass, and radius measurements for a single source will allow for new, stringent constraints on the dense-matter equation of state.
\end{abstract}

\maketitle

With the discovery and continued observations of the double pulsar system, PSR~J0737$-$3039 \cite{Lyne2004}, a neutron star moment of inertia measurement has become imminent. The moment of inertia for Pulsar A in this system can be measured from the periastron advance of the binary orbit, $\dot{\omega}$, due to relativistic spin-orbit coupling, in conjunction with the measurement of the decay of the orbital period, $\dot{P_b}$ \cite{Damour1988}. Such a measurement is expected with up to 10\% accuracy within the next five years \cite{Lyne2004, Kramer2009, Lattimer2005}.

The moment of inertia, which is a higher moment of the mass profile within the neutron star, provides a strong handle on the dense-matter equation of state (EoS). Indeed, the connection between the moment of inertia and the EoS has previously been explored. For example, \citet{Morrison2004} calculated the moment of inertia for Pulsar A for three classes of EoS and showed that a moment of inertia measurement would let us distinguish between the three classes. \citet{Bejger2005} expanded on this work and further explored the relation between the type of EoS and the resulting moment of inertia. However, both these works rely on individual equations of state at high densities and are limited by the fact that current calculations might be sampling only a restricted range of the physical possibilities.

The behavior of matter at high densities, such as those expected in a neutron star, is poorly understood. Around the nuclear saturation density, $\rho_{\rm{sat}} \simeq 2.7 \times 10^{14}$ g~cm$^{-3}$, it is possible to formulate the interactions by using an expansion in terms of few-body potentials. These potentials are then constrained by low density laboratory experiments, such as those using nucleon-nucleon scattering and the properties of light nuclei \cite{Akmal1998, Morales2002}. Other experimental constraints that provide useful input at these densities include the neutron skin thickness of heavy nuclei \cite{Centelles2009}, giant dipole resonances \cite{Trippa2008, Tamii2011, Piekarewicz2012}, and heavy ion collisions \cite{Tsang2009}. All these can be used to constrain the density-dependence of the isospin symmetry energy and, in turn, the EoS at low-densities. However, for  $\rho > \rho_{\rm{sat}}$, experimental data become sparse and the forces between particles can no longer be expanded in terms of static, few-body potentials. Moreover, quark degrees of freedom may become excited, and pion \cite{Pandharipande1975} or kaon condensates \cite{Kaplan1986} may form. 

Neutron star studies and, in particular, measurements of their moments of inertia, more directly constrain the stellar structure and the dense matter EoS than low-energy laboratory experiments. In this letter, we show how a moment of inertia measurement, $I_A$, for Pulsar A of the double pulsar system maps directly into the neutron star structure. There are two ways to accomplish this. The first is to assume an EoS throughout the star and solve for the resulting stellar structure. However, as we will demonstrate, the currently proposed EoS already show a spread in their predictions of the moment of inertia for a given stellar radius and it is unclear, given the limited range of physics explored by the current sample of EoS, whether this spread covers the entire range of possibilities. It is therefore unclear what degree of uncertainty would be associated with a neutron star radius, given a moment of inertia measurement. 

In order to address this uncertainty, we follow here a second approach that maps the moment of inertia measurements to neutron star structure in a more robust way. We employ a method that assumes an EoS only up to the nuclear saturation density and then configures the remaining mass to either maximize or minimize the moment of inertia. This method is independent of assumptions of the behavior of matter at densities above the nuclear saturation density and, thus, provides the most model-independent constraints on the neutron star structure. We show how even a weak upper bound on $I_A$ places an upper limit on the radius of Pulsar A, which will ultimately provide more stringent constraints on the EoS. Once a more precise measurement of $I_A$ is made, we show that we will be able to constrain the radius to within $\pm$1 km.

{\em{Neutron star moments of inertia.\/}} We start by showing the widely varying moments of inertia and radii that are predicted for a given neutron star mass, if different EoS are assumed throughout the star. To calculate the moment of inertia predicted by an EoS, we numerically integrate the Tolman-Oppenheimer-Volkoff (TOV) equations for stellar structure simultaneously with a relativistic version of the differential equation for the moment of inertia \cite{Glendenning1996}.

The TOV equations determine the pressure, $P$, and the enclosed mass, $M$, of the star as a function of radius, such that
\begin{equation}
\frac{\text{d}P}{\text{d}r} = - G \frac{(\rho + P/c^2)(M + 4\pi r^3 P/c^2)}{r^2 - \frac{2GMr}{c^2}}
\label{eq:dpdr}
\end{equation}
and
\begin{equation}
\frac{\text{d}M}{\text{d}r} = 4 \pi r^2 \rho,
\label{eq:dmdr}
\end{equation}
where $\rho$ is the energy density at an interior radius, $r$, and
\begin{equation}
\frac{\text{d}\nu}{\text{d}r} = \frac{2 G}{c^2} \frac{M + 4\pi r^3 P/c^2}{r^2 (1-2 \frac{G M}{rc^2})},
\label{eq:dnudr}
\end{equation}
where $e^{-\nu}$ is the $g_{tt}$ component of the metric for a slowly rotating star. The spin frequency of Pulsar A is 44 Hz, so any rotational deformations of the star will indeed be small. Equation~(\ref{eq:dnudr}) has the boundary condition $\nu(R_{\rm{NS}}) = \text{ln} \left(1 - 2 G M_{\rm{NS}}/ R_{\rm{NS}}c^2 \right)$, where $M_{\rm{NS}}$ and $R_{\rm{NS}}$ are the mass and radius of the whole star.

The additional differential equations for the moment of inertia are
\begin{equation}
\frac{\text{d}I}{\text{d}r} = \frac{8 \pi}{3} \left( \rho + \frac{P}{c^2} \right) \frac{f j r^4}{1-2\frac{GM}{rc^2}},
\label{eq:didr}
\end{equation}
and
\begin{equation}
\frac{\text{d}}{\text{d}r} \left(r^4 j \frac{\text{d}f}{\text{d}r} \right) + 4 r^3 \frac{\text{d}j}{\text{d}r} f = 0,
\label{eq:secondord}
\end{equation}
where
$f(r) \equiv 1 - \frac{\omega(r)}{\Omega}$, $j \equiv e^{-\nu/2} \left(1 - 2 \frac{GM}{rc^2} \right)^{1/2}$, $\omega(r)$ is the rotational frequency of the local inertial frame at radius $r$, and $\Omega$ is the spin frequency of the star. Equation~(\ref{eq:secondord}) is a second-order partial differential equation with the two boundary conditions
\begin{equation}
\left[ \frac{df}{dr} \right]_{r=0} = 0
\label{eq:dfdrbc}
\end{equation}
and 
\begin{equation}
f(r = R_{\rm{NS}} ) = 1 - 2 \frac{I}{R^3_{\rm{NS}}}.
\label{eq:fbc}
\end{equation}
To solve these coupled equations, we integrate Eqs.~(\ref{eq:didr}) and (\ref{eq:secondord})  outwards from the center of the star, using Eq.~(\ref{eq:dfdrbc}) as one boundary condition, and iterate it to find the value of $f_0$ for which Eq.~(\ref{eq:fbc}) is valid.

The final component necessary to integrate these equations is a relation showing how the density depends on the pressure; that is to say, we need some knowledge of the neutron star equation of state. We first compiled a large number of EoS incorporating a variety of different physics and calculation methods, as in Refs. \cite{Cook1994, Read2009}. They include purely nucleonic equations of state, such as: relativistic (BPAL12, ENG, and MPA1) and nonrelativistic (BBB2) Brueckner-Hartree-Fock EoS; variational-method EoS (e.g. FPS and WFF1-3); and a potential-method EoS (SLY). Our sampling also includes models which incorporate hyperons, pion and kaon condensates, and quarks, including, for example: a neutron-only EoS with pion condensate (PS); relativistic mean-field theory EoS with hyperons (GNH3 and H1-3); and an effective-potential EoS with hyperons (BGN1H1).

\begin{figure}[ht]
\includegraphics[width=0.5\textwidth]{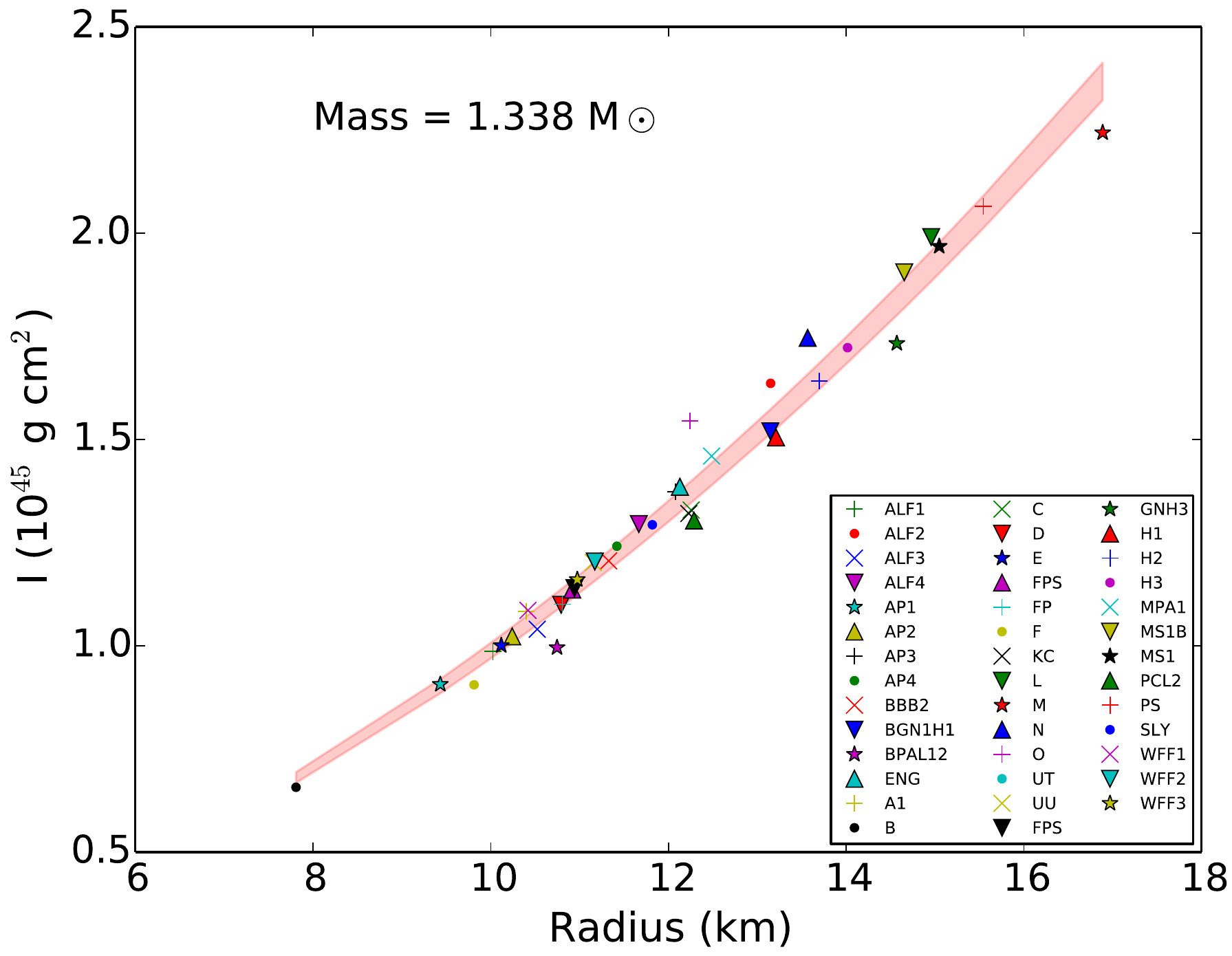}
\caption{\label{fig:EoSI} Radii and moments of inertia predicted by 41 different equations of state for a neutron star of mass $M_A = 1.338 M_{\odot}$. The red shaded region is the approximate range of moments of inertia by \citet{Lattimer2005} for EoS that do not show extreme softening at supranuclear densities. Even though their approximation follows the general trend, it does not span the entire range of radii that correspond to a given value of the moment of inertia.}
\centering
\end{figure}

For each EoS in our list, a given central density results in a unique mass and radius, as well as a moment of inertia. From these results, we choose the central density so that $M_{\rm{NS}} = 1.338 M_{\odot}$, i.e., the mass of Pulsar A. The corresponding radius and moment of inertia for each EoS are shown in Fig.~\ref{fig:EoSI}.

The moments of inertia in Fig.~\ref{fig:EoSI} vary by more than a factor of $\sim$3 and correspond to radii that vary by nearly 10 km. Figure \ref{fig:EoSI} also shows the empirical relation for moments of inertia,
\begin{equation}
I \simeq (0.237 \pm 0.008) M R^2 \left[1+ 4.2 \frac{M \text{km}}{M_{\odot} R} + 90 \left(\frac{M \text{km}}{M_{\odot} R} \right)^4 \right],
\label{eq:lattrange}
\end{equation}
obtained in Ref. \cite{Lattimer2005} by fitting to a sample of EoS which do not show significant softening at supranuclear densities and which are not self-bound. In principle, this fit could provide tight constraints on $R$ given a measurement of the moment of inertia. However, the range of neutron star radii that correspond to a given value of the moment of inertia is limited by the sample selection of EoS; it remains possible that the true neutron star EoS has not yet been formulated and is not within the uncertainties in this fit.

{\em{Absolute bounds on the moment of inertia.\/}} As shown in Fig.~\ref{fig:EoSI}, the various EoS differ by a large degree in their predictions for the moment of inertia and the corresponding radius. We present here a less model-dependent method of determining the neutron star structure from a future moment of inertia measurement. It is well known that the various EoS agree well up to $\rho \sim \rho_{\rm{sat}}$, in the regime where there is experimental data to constrain the models. Our goal is to determine the bounds that can be placed on the moment of inertia without assuming further knowledge of any EoS.

To accomplish this, we followed the formalism of \citet{Sabbadini1977}. We assumed an EoS only in the outer layer of the star, at densities below some fiducial density, $\rho_{0}$. Interior to $\rho_{0}$, we assumed one of two configurations to either maximize or minimize the resulting moment of inertia. 

\begin{figure}[ht]
\includegraphics[width=0.35\textwidth]{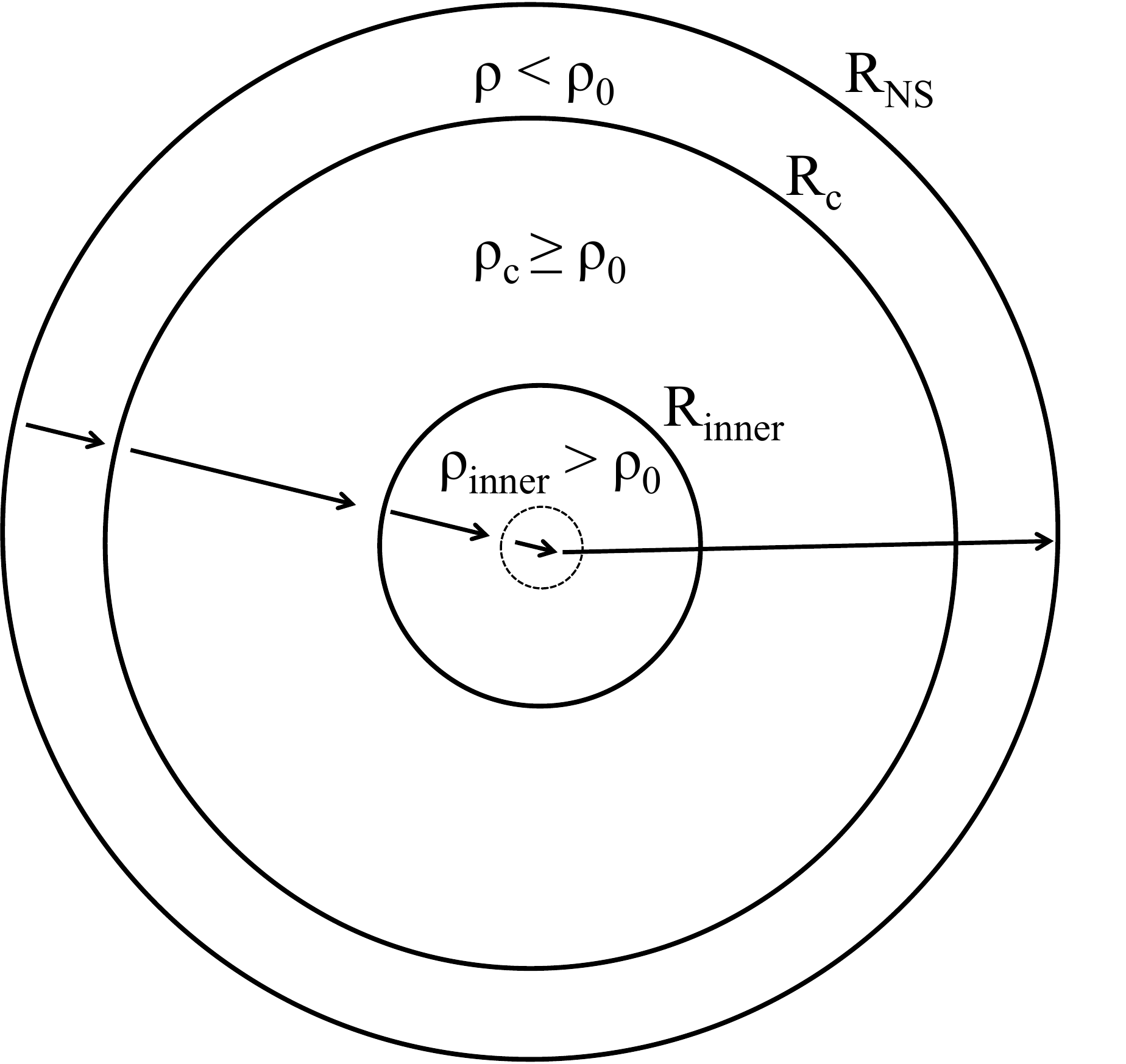}
\caption{\label{fig:minIconfig} Stellar configuration for the extremes of the moment of inertia. The outermost envelope is the region where we assume a low-density EoS. The maximum moment of inertia configuration does not have the region denoted by $\rho_{\rm{inner}}$. The minimum moment of inertia configuration requires $\rho_{\rm c} = \rho_0$.}
\centering
\end{figure}

The configuration that maximizes the moment of inertia is the one that places as much mass as possible away from the center the star while still maintaining physical stability (see Fig. \ref{fig:minIconfig}). This corresponds to a constant density core of mass $M_{\rm c}$ and radius $R_{\rm c}$, such that
\begin{equation}
\rho_{\rm{c}} = \frac{M_{\rm{c}}}{4\pi R_{\rm{c}}^3/3} \geq \rho_0.
\end{equation}
For $r < R_{\rm{c}}$, we keep the density constant, but still vary $M$ and $P$ according to the TOV equations to maintain hydrostatic equilibrium. Having determined the structure of the neutron star in this way, we then determined the moment of inertia, as described above.

We repeated this calculation starting from different stellar radii, $R_{\rm{NS}}$, but keeping the mass constant to $M_{\rm{NS}}= 1.338 M_{\odot}$. For each radius, we determined whether the resulting core is stable using the condition
\begin{equation}
\frac{4\pi}{3} R^3_{\rm c} \rho_{\rm c} \leq M_{\rm c} \leq \frac{2}{9} R_{\rm c} \left[1 - 6\pi R^2_{\rm c} P_{\rm c}+ (1 + 6\pi R^2_{\rm c} P_{\rm c})^{1/2} \right],
\label{eq:stable}
\end{equation}
(see \citep{Sabbadini1973}) which requires that the matter inside the core be a perfect fluid at all densities and that it can be described by a one-parameter EoS; that the energy density, $\rho$, is non-negative; and that both the pressure, $P$, and its derivative with respect to $\rho$, $dP/d\rho$, are non-negative.

The minimum moment of inertia configuration, on the other hand, concentrates as much mass as close to the center of the star without causing the star to collapse. This corresponds to two constant density cores, with $\rho_{\rm{c}}$ = $\rho_0$ and $\rho_{\rm{inner}} \geq \rho_{\rm{c}}$ (see Fig. \ref{fig:minIconfig}). The inner core radius, $R_{\rm{inner}}$, was determined by iteratively solving for the radius which maximizes the mass at the center of the star while still maintaining stability according to Eq.~(\ref{eq:stable}).

As discussed in Ref. \cite{Sabbadini1977}, this configuration technically requires infinite pressure at the center of the star. We assumed the pressure to be constant but finite for a small, innermost core in order to avoid numerical issues due to the infinity, and calculated the moment of inertia for the resulting stellar structure.

\begin{figure}[ht]
\includegraphics[width=0.5\textwidth]{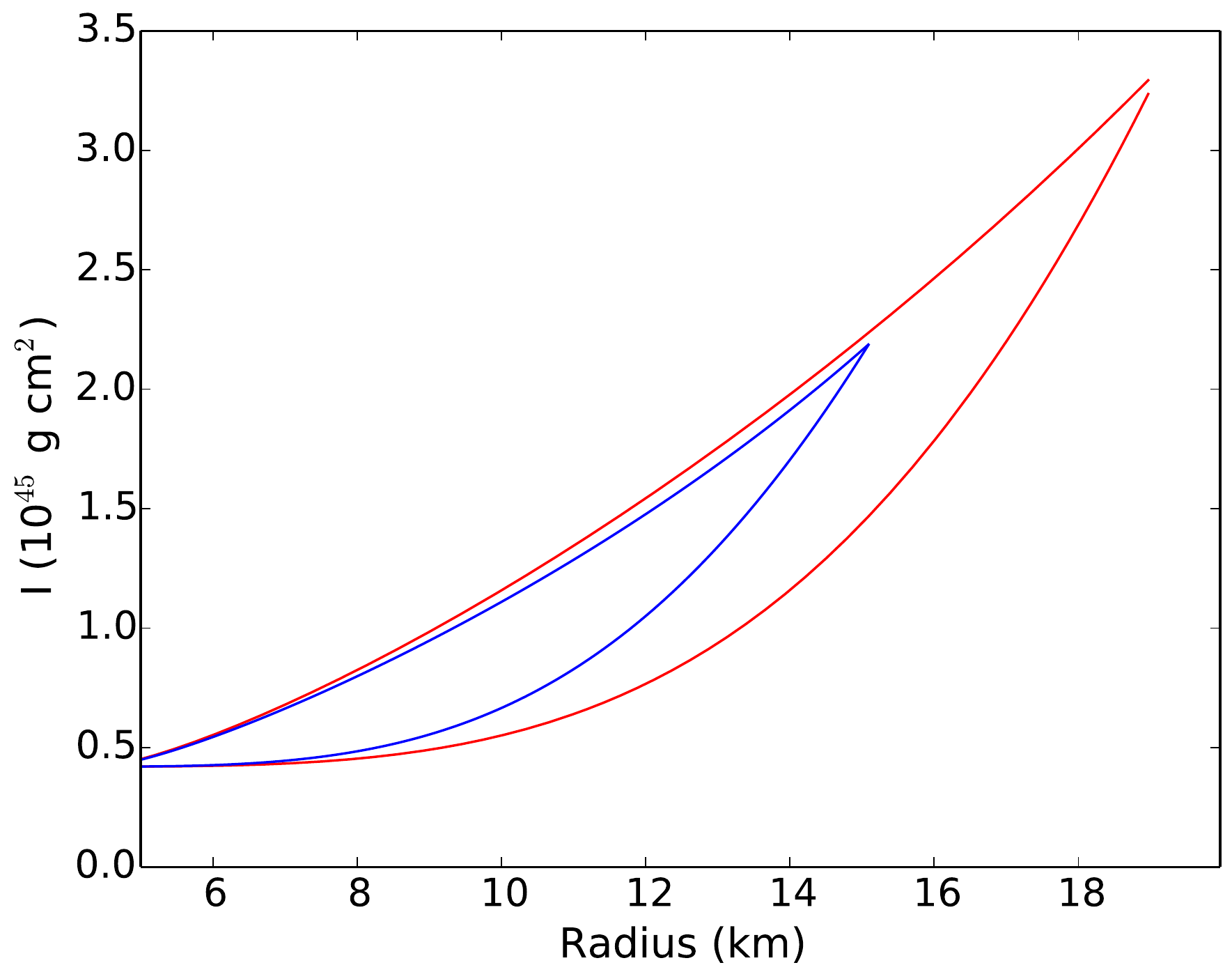}
\caption{\label{fig:Ibounds} Extreme bounds on the moment of inertia of Pulsar A as a function of its radius. The blue curve assumes the EoS, AP4, up to $\rho_0$ = $\rho_{\rm{sat}}$, while the red curve assumes AP4 up to $\rho_0$ = 0.5 $\rho_{\text{ns}}$. Interior to $\rho_0$, one of two constant-density configurations was assumed, corresponding to whether we were maximizing or minimizing the moment of inertia.}
\centering
\end{figure}

To be even more conservative, we calculated maximum and minimum moments of inertia for two cases: trusting the various EoS up to  $\rho_0$ = $\rho_{\rm{sat}}$ and trusting the EoS up to $\rho_0$ = $0.5\rho_{\rm{sat}}$. We first chose AP4 (a version of the APR equation of state) to use as a representative EoS for the integration up to $\rho=\rho_0$, as AP4 was constructed to fit low-density data \cite{Akmal1998}. We show the resulting bounds on $I_A$ in Fig.~\ref{fig:Ibounds}. 

\begin{figure}[ht]
\includegraphics[width=0.5\textwidth]{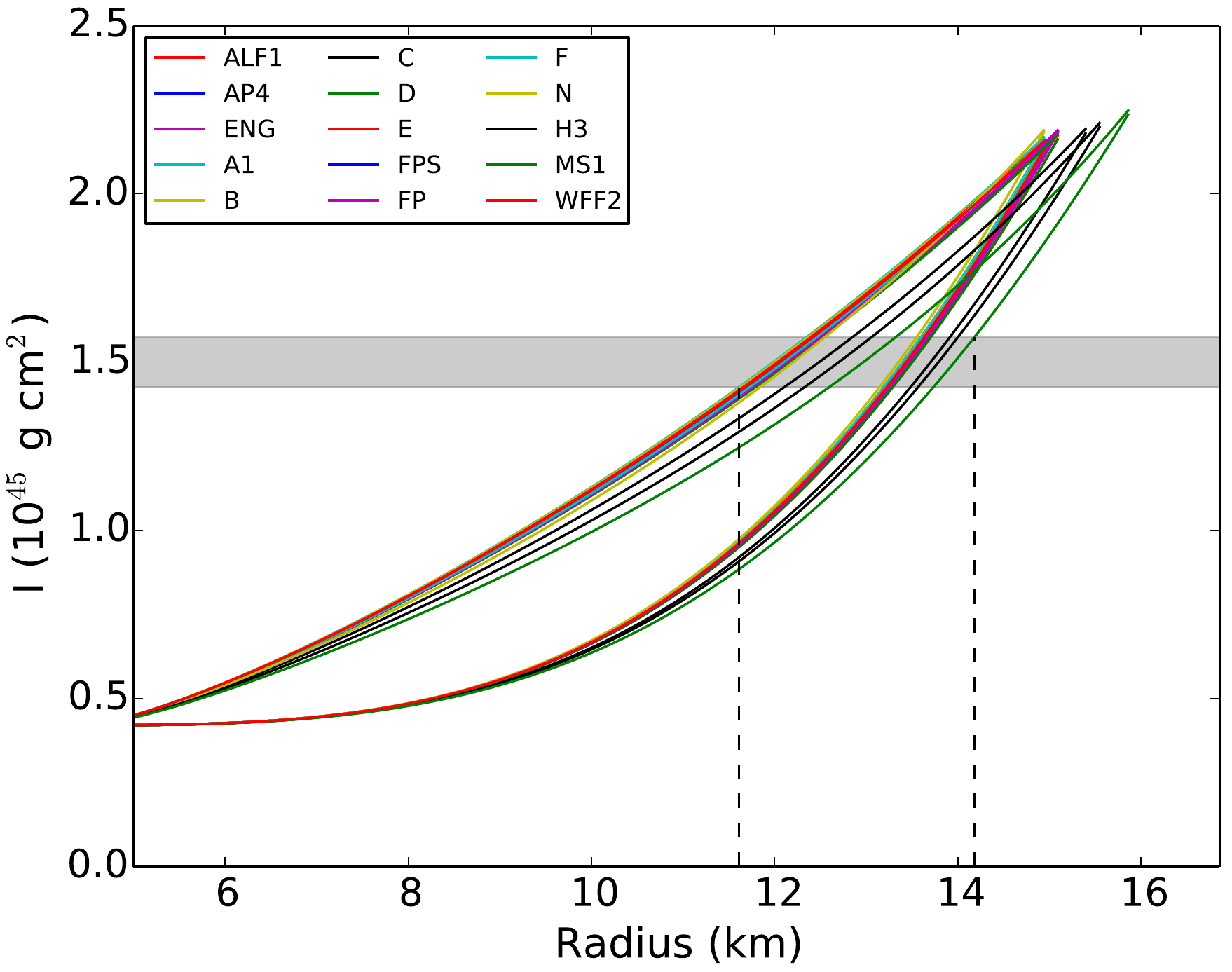}
\caption{\label{fig:Ibounds_diffEoS} Extreme bounds on the moment of inertia of Pulsar A, using different EoS for the integration up to $\rho_0$ = $\rho_{\rm{sat}}$. The shaded region represents a sample measurment of the moment of inertia to 10\% accuracy, which will lead to absolute bounds on the radius of approximately $\pm$1 km.}
\centering
\end{figure}

We then varied this assumption and used other EoS for the low-density portion of the integration, the results of which are shown in Fig.~\ref{fig:Ibounds_diffEoS}. We also show in this figure an example moment of inertia measurements with 10\% accuracy.  We show that even for the extreme configurations that are obtained for the maximum and minimum $I_A$, we constrain the radius to within $\pm$1~km. This constraint can become even tighter if additional considerations about the neutron star structure are incorporated. For example, the observation of a 2 $M_{\odot}$ neutron star already places a fairly EoS-independent lower limit on the radius by excluding radii less than 8.3 km \cite{Ozel2010a}.

Figure \ref{fig:Ibounds_diffEoS} shows that these bounds depend very weakly on the low-density EoS; i.e., assuming different EoS produces roughly the same bounds on the radius of Pulsar A, given a measurement of $I_A$. This is expected since all EoS agree fairly well with each other up to $\rho \sim \rho_{\rm{sat}}$. Therefore, a moment of inertia measurement of a pulsar of known mass will directly lead to a model-independent measurement of its radius. This is important for a direct comparison of a moment of inertia measurement to other astrophysical measurements of neutron star radii such as those from spectroscopic methods \cite{Ozel2015}, without requiring any assumptions about the EoS. It will also potentially allow measurements of neutron star radii at different masses.

As in the case of neutron star radii, the measurement of the moment of inertia will directly 
lead to quantitative constraints on the ultradense matter equation of state \cite{Lattimer2001, Read2009, Ozel2009, Lattimer2014a}.  As an example, a measurement 
of the moment of inertia with 10\% uncertainty can be directly translated into constraints on the 
magnitude $S$ and the slope $L$ of the symmetry energy at nuclear saturation 
density. Quantitatively, for a measurement of $I=1.3 \times 10^{45}$~g~cm$^2$ for Pulsar A 
with a 10\% uncertainty, the framework presented in this paper will lead to a radius inference  
with a similar level of uncertainty. This leads to a measurement of the pressure at nuclear 
saturation density of $P_{\rm sat} \sim 2.6$~MeV~fm$^{-3}$, with only somewhat larger 
uncertainty, as well as of the symmetry energy parameter $L \sim 3 P_{\rm sat}/\rho_{\rm sat} \simeq 
48 \pm 10$~MeV.

{\em{Acknowledgements.\/}} We thank Norbert Wex, Paulo Freire, and Michael Kramer for numerous useful discussions. This work was supported in part by the National Science Foundation.

\bibliography{ms_accepted}

\begin{thebibliography}{25}%
\makeatletter
\providecommand \@ifxundefined [1]{%
 \@ifx{#1\undefined}
}%
\providecommand \@ifnum [1]{%
 \ifnum #1\expandafter \@firstoftwo
 \else \expandafter \@secondoftwo
 \fi
}%
\providecommand \@ifx [1]{%
 \ifx #1\expandafter \@firstoftwo
 \else \expandafter \@secondoftwo
 \fi
}%
\providecommand \natexlab [1]{#1}%
\providecommand \enquote  [1]{``#1''}%
\providecommand \bibnamefont  [1]{#1}%
\providecommand \bibfnamefont [1]{#1}%
\providecommand \citenamefont [1]{#1}%
\providecommand \href@noop [0]{\@secondoftwo}%
\providecommand \href [0]{\begingroup \@sanitize@url \@href}%
\providecommand \@href[1]{\@@startlink{#1}\@@href}%
\providecommand \@@href[1]{\endgroup#1\@@endlink}%
\providecommand \@sanitize@url [0]{\catcode `\\12\catcode `\$12\catcode
  `\&12\catcode `\#12\catcode `\^12\catcode `\_12\catcode `\%12\relax}%
\providecommand \@@startlink[1]{}%
\providecommand \@@endlink[0]{}%
\providecommand \url  [0]{\begingroup\@sanitize@url \@url }%
\providecommand \@url [1]{\endgroup\@href {#1}{\urlprefix }}%
\providecommand \urlprefix  [0]{URL }%
\providecommand \Eprint [0]{\href }%
\providecommand \doibase [0]{http://dx.doi.org/}%
\providecommand \selectlanguage [0]{\@gobble}%
\providecommand \bibinfo  [0]{\@secondoftwo}%
\providecommand \bibfield  [0]{\@secondoftwo}%
\providecommand \translation [1]{[#1]}%
\providecommand \BibitemOpen [0]{}%
\providecommand \bibitemStop [0]{}%
\providecommand \bibitemNoStop [0]{.\EOS\space}%
\providecommand \EOS [0]{\spacefactor3000\relax}%
\providecommand \BibitemShut  [1]{\csname bibitem#1\endcsname}%
\let\auto@bib@innerbib\@empty
\bibitem [{\citenamefont {{Lyne}}\ \emph {et~al.}(2004)\citenamefont {{Lyne}},
  \citenamefont {{Burgay}}, \citenamefont {{Kramer}}, \citenamefont
  {{Possenti}}, \citenamefont {{Manchester}}, \citenamefont {{Camilo}},
  \citenamefont {{McLaughlin}}, \citenamefont {{Lorimer}}, \citenamefont
  {{D'Amico}}, \citenamefont {{Joshi}} \emph {et~al.}}]{Lyne2004}%
  \BibitemOpen
  \bibfield  {author} {\bibinfo {author} {\bibfnamefont {A.~G.}\ \bibnamefont
  {{Lyne}}}, \bibinfo {author} {\bibfnamefont {M.}~\bibnamefont {{Burgay}}},
  \bibinfo {author} {\bibfnamefont {M.}~\bibnamefont {{Kramer}}}, \bibinfo
  {author} {\bibfnamefont {A.}~\bibnamefont {{Possenti}}}, \bibinfo {author}
  {\bibfnamefont {R.~N.}\ \bibnamefont {{Manchester}}}, \bibinfo {author}
  {\bibfnamefont {F.}~\bibnamefont {{Camilo}}}, \bibinfo {author}
  {\bibfnamefont {M.~A.}\ \bibnamefont {{McLaughlin}}}, \bibinfo {author}
  {\bibfnamefont {D.~R.}\ \bibnamefont {{Lorimer}}}, \bibinfo {author}
  {\bibfnamefont {N.}~\bibnamefont {{D'Amico}}}, \bibinfo {author}
  {\bibfnamefont {B.~C.}\ \bibnamefont {{Joshi}}},  \emph {et~al.},\ }\href
  {\doibase 10.1126/science.1094645} {\bibfield  {journal} {\bibinfo  {journal}
  {Science}\ }\textbf {\bibinfo {volume} {303}},\ \bibinfo {pages} {1153}
  (\bibinfo {year} {2004})}\BibitemShut {NoStop}%
\bibitem [{\citenamefont {{Damour}}\ and\ \citenamefont
  {{Schafer}}(1988)}]{Damour1988}%
  \BibitemOpen
  \bibfield  {author} {\bibinfo {author} {\bibfnamefont {T.}~\bibnamefont
  {{Damour}}}\ and\ \bibinfo {author} {\bibfnamefont {G.}~\bibnamefont
  {{Schafer}}},\ }\href {\doibase 10.1007/BF02828697} {\bibfield  {journal}
  {\bibinfo  {journal} {Nuovo Cimento B Series}\ }\textbf {\bibinfo {volume}
  {101}},\ \bibinfo {pages} {127} (\bibinfo {year} {1988})}\BibitemShut
  {NoStop}%
\bibitem [{\citenamefont {{Kramer}}\ and\ \citenamefont
  {{Wex}}(2009)}]{Kramer2009}%
  \BibitemOpen
  \bibfield  {author} {\bibinfo {author} {\bibfnamefont {M.}~\bibnamefont
  {{Kramer}}}\ and\ \bibinfo {author} {\bibfnamefont {N.}~\bibnamefont
  {{Wex}}},\ }\href {\doibase 10.1088/0264-9381/26/7/073001} {\bibfield
  {journal} {\bibinfo  {journal} {Classical and Quantum Gravity}\ }\textbf
  {\bibinfo {volume} {26}},\ \bibinfo {eid} {073001} (\bibinfo {year}
  {2009})}\BibitemShut {NoStop}%
\bibitem [{\citenamefont {{Lattimer}}\ and\ \citenamefont
  {{Schutz}}(2005)}]{Lattimer2005}%
  \BibitemOpen
  \bibfield  {author} {\bibinfo {author} {\bibfnamefont {J.~M.}\ \bibnamefont
  {{Lattimer}}}\ and\ \bibinfo {author} {\bibfnamefont {B.~F.}\ \bibnamefont
  {{Schutz}}},\ }\href {\doibase 10.1086/431543} {\bibfield  {journal}
  {\bibinfo  {journal} {\apj}\ }\textbf {\bibinfo {volume} {629}},\ \bibinfo
  {pages} {979} (\bibinfo {year} {2005})}\BibitemShut {NoStop}%
\bibitem [{\citenamefont {{Morrison}}\ \emph {et~al.}(2004)\citenamefont
  {{Morrison}}, \citenamefont {{Baumgarte}}, \citenamefont {{Shapiro}},\ and\
  \citenamefont {{Pandharipande}}}]{Morrison2004}%
  \BibitemOpen
  \bibfield  {author} {\bibinfo {author} {\bibfnamefont {I.~A.}\ \bibnamefont
  {{Morrison}}}, \bibinfo {author} {\bibfnamefont {T.~W.}\ \bibnamefont
  {{Baumgarte}}}, \bibinfo {author} {\bibfnamefont {S.~L.}\ \bibnamefont
  {{Shapiro}}}, \ and\ \bibinfo {author} {\bibfnamefont {V.~R.}\ \bibnamefont
  {{Pandharipande}}},\ }\href {\doibase 10.1086/427235} {\bibfield  {journal}
  {\bibinfo  {journal} {The Astrophysical Journal Letters}\ }\textbf {\bibinfo
  {volume} {617}},\ \bibinfo {pages} {L135} (\bibinfo {year}
  {2004})}\BibitemShut {NoStop}%
\bibitem [{\citenamefont {{Bejger}}\ \emph {et~al.}(2005)\citenamefont
  {{Bejger}}, \citenamefont {{Bulik}},\ and\ \citenamefont
  {{Haensel}}}]{Bejger2005}%
  \BibitemOpen
  \bibfield  {author} {\bibinfo {author} {\bibfnamefont {M.}~\bibnamefont
  {{Bejger}}}, \bibinfo {author} {\bibfnamefont {T.}~\bibnamefont {{Bulik}}}, \
  and\ \bibinfo {author} {\bibfnamefont {P.}~\bibnamefont {{Haensel}}},\ }\href
  {\doibase 10.1111/j.1365-2966.2005.09575.x} {\bibfield  {journal} {\bibinfo
  {journal} {Monthly Notices of the Royal Astronomical Society}\ }\textbf
  {\bibinfo {volume} {364}},\ \bibinfo {pages} {635} (\bibinfo {year}
  {2005})}\BibitemShut {NoStop}%
\bibitem [{\citenamefont {{Akmal}}\ \emph {et~al.}(1998)\citenamefont
  {{Akmal}}, \citenamefont {{Pandharipande}},\ and\ \citenamefont
  {{Ravenhall}}}]{Akmal1998}%
  \BibitemOpen
  \bibfield  {author} {\bibinfo {author} {\bibfnamefont {A.}~\bibnamefont
  {{Akmal}}}, \bibinfo {author} {\bibfnamefont {V.~R.}\ \bibnamefont
  {{Pandharipande}}}, \ and\ \bibinfo {author} {\bibfnamefont {D.~G.}\
  \bibnamefont {{Ravenhall}}},\ }\href {\doibase 10.1103/PhysRevC.58.1804}
  {\bibfield  {journal} {\bibinfo  {journal} {\prc}\ }\textbf {\bibinfo
  {volume} {58}},\ \bibinfo {pages} {1804} (\bibinfo {year}
  {1998})}\BibitemShut {NoStop}%
\bibitem [{\citenamefont {{Morales}}\ \emph {et~al.}(2002)\citenamefont
  {{Morales}}, \citenamefont {{Pandharipande}},\ and\ \citenamefont
  {{Ravenhall}}}]{Morales2002}%
  \BibitemOpen
  \bibfield  {author} {\bibinfo {author} {\bibfnamefont {J.}~\bibnamefont
  {{Morales}}}, \bibinfo {author} {\bibfnamefont {V.~R.}\ \bibnamefont
  {{Pandharipande}}}, \ and\ \bibinfo {author} {\bibfnamefont {D.~G.}\
  \bibnamefont {{Ravenhall}}},\ }\href {\doibase 10.1103/PhysRevC.66.054308}
  {\bibfield  {journal} {\bibinfo  {journal} {\prc}\ }\textbf {\bibinfo
  {volume} {66}},\ \bibinfo {eid} {054308} (\bibinfo {year}
  {2002})}\BibitemShut {NoStop}%
\bibitem [{\citenamefont {{Centelles}}\ \emph {et~al.}(2009)\citenamefont
  {{Centelles}}, \citenamefont {{Roca-Maza}}, \citenamefont {{Vi{\~n}as}},\
  and\ \citenamefont {{Warda}}}]{Centelles2009}%
  \BibitemOpen
  \bibfield  {author} {\bibinfo {author} {\bibfnamefont {M.}~\bibnamefont
  {{Centelles}}}, \bibinfo {author} {\bibfnamefont {X.}~\bibnamefont
  {{Roca-Maza}}}, \bibinfo {author} {\bibfnamefont {X.}~\bibnamefont
  {{Vi{\~n}as}}}, \ and\ \bibinfo {author} {\bibfnamefont {M.}~\bibnamefont
  {{Warda}}},\ }\href {\doibase 10.1103/PhysRevLett.102.122502} {\bibfield
  {journal} {\bibinfo  {journal} {Physical Review Letters}\ }\textbf {\bibinfo
  {volume} {102}},\ \bibinfo {eid} {122502} (\bibinfo {year}
  {2009})}\BibitemShut {NoStop}%
\bibitem [{\citenamefont {{Trippa}}\ \emph {et~al.}(2008)\citenamefont
  {{Trippa}}, \citenamefont {{Col{\`o}}},\ and\ \citenamefont
  {{Vigezzi}}}]{Trippa2008}%
  \BibitemOpen
  \bibfield  {author} {\bibinfo {author} {\bibfnamefont {L.}~\bibnamefont
  {{Trippa}}}, \bibinfo {author} {\bibfnamefont {G.}~\bibnamefont
  {{Col{\`o}}}}, \ and\ \bibinfo {author} {\bibfnamefont {E.}~\bibnamefont
  {{Vigezzi}}},\ }\href {\doibase 10.1103/PhysRevC.77.061304} {\bibfield
  {journal} {\bibinfo  {journal} {\prc}\ }\textbf {\bibinfo {volume} {77}},\
  \bibinfo {eid} {061304} (\bibinfo {year} {2008})}\BibitemShut {NoStop}%
\bibitem [{\citenamefont {{Tamii}}\ \emph {et~al.}(2011)\citenamefont
  {{Tamii}}, \citenamefont {{Poltoratska}}, \citenamefont {{von
  Neumann-Cosel}}, \citenamefont {{Fujita}}, \citenamefont {{Adachi}},
  \citenamefont {{Bertulani}}, \citenamefont {{Carter}}, \citenamefont
  {{Dozono}}, \citenamefont {{Fujita}}, \citenamefont {{Fujita}} \emph
  {et~al.}}]{Tamii2011}%
  \BibitemOpen
  \bibfield  {author} {\bibinfo {author} {\bibfnamefont {A.}~\bibnamefont
  {{Tamii}}}, \bibinfo {author} {\bibfnamefont {I.}~\bibnamefont
  {{Poltoratska}}}, \bibinfo {author} {\bibfnamefont {P.}~\bibnamefont {{von
  Neumann-Cosel}}}, \bibinfo {author} {\bibfnamefont {Y.}~\bibnamefont
  {{Fujita}}}, \bibinfo {author} {\bibfnamefont {T.}~\bibnamefont {{Adachi}}},
  \bibinfo {author} {\bibfnamefont {C.~A.}\ \bibnamefont {{Bertulani}}},
  \bibinfo {author} {\bibfnamefont {J.}~\bibnamefont {{Carter}}}, \bibinfo
  {author} {\bibfnamefont {M.}~\bibnamefont {{Dozono}}}, \bibinfo {author}
  {\bibfnamefont {H.}~\bibnamefont {{Fujita}}}, \bibinfo {author}
  {\bibfnamefont {K.}~\bibnamefont {{Fujita}}},  \emph {et~al.},\ }\href
  {\doibase 10.1103/PhysRevLett.107.062502} {\bibfield  {journal} {\bibinfo
  {journal} {Physical Review Letters}\ }\textbf {\bibinfo {volume} {107}},\
  \bibinfo {eid} {062502} (\bibinfo {year} {2011})}\BibitemShut {NoStop}%
\bibitem [{\citenamefont {{Piekarewicz}}\ \emph {et~al.}(2012)\citenamefont
  {{Piekarewicz}}, \citenamefont {{Agrawal}}, \citenamefont {{Col{\`o}}},
  \citenamefont {{Nazarewicz}}, \citenamefont {{Paar}}, \citenamefont
  {{Reinhard}}, \citenamefont {{Roca-Maza}},\ and\ \citenamefont
  {{Vretenar}}}]{Piekarewicz2012}%
  \BibitemOpen
  \bibfield  {author} {\bibinfo {author} {\bibfnamefont {J.}~\bibnamefont
  {{Piekarewicz}}}, \bibinfo {author} {\bibfnamefont {B.~K.}\ \bibnamefont
  {{Agrawal}}}, \bibinfo {author} {\bibfnamefont {G.}~\bibnamefont
  {{Col{\`o}}}}, \bibinfo {author} {\bibfnamefont {W.}~\bibnamefont
  {{Nazarewicz}}}, \bibinfo {author} {\bibfnamefont {N.}~\bibnamefont
  {{Paar}}}, \bibinfo {author} {\bibfnamefont {P.-G.}\ \bibnamefont
  {{Reinhard}}}, \bibinfo {author} {\bibfnamefont {X.}~\bibnamefont
  {{Roca-Maza}}}, \ and\ \bibinfo {author} {\bibfnamefont {D.}~\bibnamefont
  {{Vretenar}}},\ }\href {\doibase 10.1103/PhysRevC.85.041302} {\bibfield
  {journal} {\bibinfo  {journal} {\prc}\ }\textbf {\bibinfo {volume} {85}},\
  \bibinfo {eid} {041302} (\bibinfo {year} {2012})}\BibitemShut {NoStop}%
\bibitem [{\citenamefont {{Tsang}}\ \emph {et~al.}(2009)\citenamefont
  {{Tsang}}, \citenamefont {{Zhang}}, \citenamefont {{Danielewicz}},
  \citenamefont {{Famiano}}, \citenamefont {{Li}}, \citenamefont {{Lynch}},\
  and\ \citenamefont {{Steiner}}}]{Tsang2009}%
  \BibitemOpen
  \bibfield  {author} {\bibinfo {author} {\bibfnamefont {M.~B.}\ \bibnamefont
  {{Tsang}}}, \bibinfo {author} {\bibfnamefont {Y.}~\bibnamefont {{Zhang}}},
  \bibinfo {author} {\bibfnamefont {P.}~\bibnamefont {{Danielewicz}}}, \bibinfo
  {author} {\bibfnamefont {M.}~\bibnamefont {{Famiano}}}, \bibinfo {author}
  {\bibfnamefont {Z.}~\bibnamefont {{Li}}}, \bibinfo {author} {\bibfnamefont
  {W.~G.}\ \bibnamefont {{Lynch}}}, \ and\ \bibinfo {author} {\bibfnamefont
  {A.~W.}\ \bibnamefont {{Steiner}}},\ }\href {\doibase
  10.1103/PhysRevLett.102.122701} {\bibfield  {journal} {\bibinfo  {journal}
  {Physical Review Letters}\ }\textbf {\bibinfo {volume} {102}},\ \bibinfo
  {eid} {122701} (\bibinfo {year} {2009})}\BibitemShut {NoStop}%
\bibitem [{\citenamefont {{Pandharipande}}\ and\ \citenamefont
  {{Smith}}(1975)}]{Pandharipande1975}%
  \BibitemOpen
  \bibfield  {author} {\bibinfo {author} {\bibfnamefont {V.~R.}\ \bibnamefont
  {{Pandharipande}}}\ and\ \bibinfo {author} {\bibfnamefont {R.~A.}\
  \bibnamefont {{Smith}}},\ }\href {\doibase 10.1016/0375-9474(75)90415-7}
  {\bibfield  {journal} {\bibinfo  {journal} {Nuclear Physics A}\ }\textbf
  {\bibinfo {volume} {237}},\ \bibinfo {pages} {507} (\bibinfo {year}
  {1975})}\BibitemShut {NoStop}%
\bibitem [{\citenamefont {{Kaplan}}\ and\ \citenamefont
  {{Nelson}}(1986)}]{Kaplan1986}%
  \BibitemOpen
  \bibfield  {author} {\bibinfo {author} {\bibfnamefont {D.~B.}\ \bibnamefont
  {{Kaplan}}}\ and\ \bibinfo {author} {\bibfnamefont {A.~E.}\ \bibnamefont
  {{Nelson}}},\ }\href {\doibase 10.1016/0370-2693(86)90331-X} {\bibfield
  {journal} {\bibinfo  {journal} {Physics Letters B}\ }\textbf {\bibinfo
  {volume} {175}},\ \bibinfo {pages} {57} (\bibinfo {year} {1986})}\BibitemShut
  {NoStop}%
\bibitem [{\citenamefont {{Glendenning}}(1996)}]{Glendenning1996}%
  \BibitemOpen
  \bibfield  {author} {\bibinfo {author} {\bibfnamefont {N.}~\bibnamefont
  {{Glendenning}}},\ }\href@noop {} {\emph {\bibinfo {title} {Compact Stars.~
  Nuclear Physics, Particle Physics and General Relativity, Approx.~390 pp.~90
  figs..~Springer-Verlag New York.~ Also Astronomy and Astrophysics Library}}}\
  (\bibinfo {year} {1996})\BibitemShut {NoStop}%
\bibitem [{\citenamefont {{Cook}}\ \emph {et~al.}(1994)\citenamefont {{Cook}},
  \citenamefont {{Shapiro}},\ and\ \citenamefont {{Teukolsky}}}]{Cook1994}%
  \BibitemOpen
  \bibfield  {author} {\bibinfo {author} {\bibfnamefont {G.~B.}\ \bibnamefont
  {{Cook}}}, \bibinfo {author} {\bibfnamefont {S.~L.}\ \bibnamefont
  {{Shapiro}}}, \ and\ \bibinfo {author} {\bibfnamefont {S.~A.}\ \bibnamefont
  {{Teukolsky}}},\ }\href {\doibase 10.1086/173934} {\bibfield  {journal}
  {\bibinfo  {journal} {\apj}\ }\textbf {\bibinfo {volume} {424}},\ \bibinfo
  {pages} {823} (\bibinfo {year} {1994})}\BibitemShut {NoStop}%
\bibitem [{\citenamefont {{Read}}\ \emph {et~al.}(2009)\citenamefont {{Read}},
  \citenamefont {{Lackey}}, \citenamefont {{Owen}},\ and\ \citenamefont
  {{Friedman}}}]{Read2009}%
  \BibitemOpen
  \bibfield  {author} {\bibinfo {author} {\bibfnamefont {J.~S.}\ \bibnamefont
  {{Read}}}, \bibinfo {author} {\bibfnamefont {B.~D.}\ \bibnamefont
  {{Lackey}}}, \bibinfo {author} {\bibfnamefont {B.~J.}\ \bibnamefont
  {{Owen}}}, \ and\ \bibinfo {author} {\bibfnamefont {J.~L.}\ \bibnamefont
  {{Friedman}}},\ }\href {\doibase 10.1103/PhysRevD.79.124032} {\bibfield
  {journal} {\bibinfo  {journal} {\prd}\ }\textbf {\bibinfo {volume} {79}},\
  \bibinfo {eid} {124032} (\bibinfo {year} {2009})}\BibitemShut {NoStop}%
\bibitem [{\citenamefont {{Sabbadini}}\ and\ \citenamefont
  {{Hartle}}(1977)}]{Sabbadini1977}%
  \BibitemOpen
  \bibfield  {author} {\bibinfo {author} {\bibfnamefont {A.~G.}\ \bibnamefont
  {{Sabbadini}}}\ and\ \bibinfo {author} {\bibfnamefont {J.~B.}\ \bibnamefont
  {{Hartle}}},\ }\href {\doibase 10.1016/0003-4916(77)90047-1} {\bibfield
  {journal} {\bibinfo  {journal} {Annals of Physics}\ }\textbf {\bibinfo
  {volume} {104}},\ \bibinfo {pages} {95} (\bibinfo {year} {1977})}\BibitemShut
  {NoStop}%
\bibitem [{\citenamefont {{Sabbadini}}\ and\ \citenamefont
  {{Hartle}}(1973)}]{Sabbadini1973}%
  \BibitemOpen
  \bibfield  {author} {\bibinfo {author} {\bibfnamefont {A.~G.}\ \bibnamefont
  {{Sabbadini}}}\ and\ \bibinfo {author} {\bibfnamefont {J.~B.}\ \bibnamefont
  {{Hartle}}},\ }\href {\doibase 10.1007/BF00648231} {\bibfield  {journal}
  {\bibinfo  {journal} {Astrophysics and Space Science}\ }\textbf {\bibinfo
  {volume} {25}},\ \bibinfo {pages} {117} (\bibinfo {year} {1973})}\BibitemShut
  {NoStop}%
\bibitem [{\citenamefont {{{\"O}zel}}\ \emph {et~al.}(2010)\citenamefont
  {{{\"O}zel}}, \citenamefont {{Psaltis}}, \citenamefont {{Ransom}},
  \citenamefont {{Demorest}},\ and\ \citenamefont {{Alford}}}]{Ozel2010a}%
  \BibitemOpen
  \bibfield  {author} {\bibinfo {author} {\bibfnamefont {F.}~\bibnamefont
  {{{\"O}zel}}}, \bibinfo {author} {\bibfnamefont {D.}~\bibnamefont
  {{Psaltis}}}, \bibinfo {author} {\bibfnamefont {S.}~\bibnamefont {{Ransom}}},
  \bibinfo {author} {\bibfnamefont {P.}~\bibnamefont {{Demorest}}}, \ and\
  \bibinfo {author} {\bibfnamefont {M.}~\bibnamefont {{Alford}}},\ }\href
  {\doibase 10.1088/2041-8205/724/2/L199} {\bibfield  {journal} {\bibinfo
  {journal} {Astrophys. J. Letters}\ }\textbf {\bibinfo {volume} {724}},\
  \bibinfo {pages} {L199} (\bibinfo {year} {2010})}\BibitemShut {NoStop}%
\bibitem [{\citenamefont {{Ozel}}\ \emph {et~al.}(2015)\citenamefont {{Ozel}},
  \citenamefont {{Psaltis}}, \citenamefont {{Guver}}, \citenamefont {{Baym}},
  \citenamefont {{Heinke}},\ and\ \citenamefont {{Guillot}}}]{Ozel2015}%
  \BibitemOpen
  \bibfield  {author} {\bibinfo {author} {\bibfnamefont {F.}~\bibnamefont
  {{Ozel}}}, \bibinfo {author} {\bibfnamefont {D.}~\bibnamefont {{Psaltis}}},
  \bibinfo {author} {\bibfnamefont {T.}~\bibnamefont {{Guver}}}, \bibinfo
  {author} {\bibfnamefont {G.}~\bibnamefont {{Baym}}}, \bibinfo {author}
  {\bibfnamefont {C.}~\bibnamefont {{Heinke}}}, \ and\ \bibinfo {author}
  {\bibfnamefont {S.}~\bibnamefont {{Guillot}}},\ }\href@noop {} {\bibfield
  {journal} {\bibinfo  {journal} {ArXiv e-prints}\ } (\bibinfo {year}
  {2015})},\ \Eprint {http://arxiv.org/abs/1505.05155} {arXiv:1505.05155
  [astro-ph.HE]} \BibitemShut {NoStop}%
\bibitem [{\citenamefont {{Lattimer}}\ and\ \citenamefont
  {{Prakash}}(2001)}]{Lattimer2001}%
  \BibitemOpen
  \bibfield  {author} {\bibinfo {author} {\bibfnamefont {J.~M.}\ \bibnamefont
  {{Lattimer}}}\ and\ \bibinfo {author} {\bibfnamefont {M.}~\bibnamefont
  {{Prakash}}},\ }\href {\doibase 10.1086/319702} {\bibfield  {journal}
  {\bibinfo  {journal} {\apj}\ }\textbf {\bibinfo {volume} {550}},\ \bibinfo
  {pages} {426} (\bibinfo {year} {2001})},\ \Eprint
  {http://arxiv.org/abs/astro-ph/0002232} {astro-ph/0002232} \BibitemShut
  {NoStop}%
\bibitem [{\citenamefont {{{\"O}zel}}\ and\ \citenamefont
  {{Psaltis}}(2009)}]{Ozel2009}%
  \BibitemOpen
  \bibfield  {author} {\bibinfo {author} {\bibfnamefont {F.}~\bibnamefont
  {{{\"O}zel}}}\ and\ \bibinfo {author} {\bibfnamefont {D.}~\bibnamefont
  {{Psaltis}}},\ }\href {\doibase 10.1103/PhysRevD.80.103003} {\bibfield
  {journal} {\bibinfo  {journal} {\prd}\ }\textbf {\bibinfo {volume} {80}},\
  \bibinfo {eid} {103003} (\bibinfo {year} {2009})},\ \Eprint
  {http://arxiv.org/abs/0905.1959} {arXiv:0905.1959 [astro-ph.HE]} \BibitemShut
  {NoStop}%
\bibitem [{\citenamefont {{Lattimer}}\ and\ \citenamefont
  {{Steiner}}(2014)}]{Lattimer2014a}%
  \BibitemOpen
  \bibfield  {author} {\bibinfo {author} {\bibfnamefont {J.~M.}\ \bibnamefont
  {{Lattimer}}}\ and\ \bibinfo {author} {\bibfnamefont {A.~W.}\ \bibnamefont
  {{Steiner}}},\ }\href {\doibase 10.1140/epja/i2014-14040-y} {\bibfield
  {journal} {\bibinfo  {journal} {European Physical Journal A}\ }\textbf
  {\bibinfo {volume} {50}},\ \bibinfo {eid} {40} (\bibinfo {year} {2014})},\
  \Eprint {http://arxiv.org/abs/1403.1186} {arXiv:1403.1186 [nucl-th]}
  \BibitemShut {NoStop}%
\end{thebibliography}%
\bibliographystyle{apsrev4-1}

\end{document}